\documentclass[12pt]{article}
\usepackage{amssymb,graphicx,times}
\newcommand{\de}{\partial}

\begin{document}

\thispagestyle{empty}
\setcounter{page}{1}

\vspace*{0.88truein}
\centerline{\bf GROWTH-OPTIMAL STRATEGIES WITH QUADRATIC FRICTION}
\vspace*{0.087truein}
\centerline{\bf OVER FINITE-TIME INVESTMENT HORIZONS}
\vspace*{0.37truein}
\vspace*{0.37truein}
\baselineskip=10pt
\centerline{\footnotesize ERIK AURELL}
\vspace*{0.015truein}
\centerline{\footnotesize\it Swedish Institute of Computer Science}
\baselineskip=10pt
\centerline{\footnotesize\it SICS Box 1263 SE-16429, Kista, Sweden}
\baselineskip=10pt
\centerline{\footnotesize\it erik.aurell@sics.se}
\centerline{\footnotesize\it and}
\baselineskip=10pt
\centerline{\footnotesize\it Department of Numerical Analyis and Computer Science}
\baselineskip=10pt
\centerline{\footnotesize\it KTH-Royal Institute of Technology,  SE-100 44 Stockholm, Sweden}
\vspace*{0.37truein}
\centerline{\footnotesize PAOLO MURATORE-GINANNESCHI}
\vspace*{0.015truein}
\centerline{\footnotesize\it Departments of Mathematics}
\baselineskip=10pt
\centerline{\footnotesize\it P.O. Box 4 FIN-00014, University of Helsinki, Finland}
\baselineskip=10pt
\centerline{\footnotesize\it paolo.muratore-ginanneschi@helsinki.fi}
\baselineskip=10pt
\vspace*{10pt}

\begin{abstract}We investigate the growth optimal strategy over a finite
time horizon for a stock and bond portfolio in an 
analytically solvable multiplicative Markovian market model. We show that 
the optimal strategy consists in holding the amount of capital invested in stocks 
within an interval around an ideal optimal investment.
The size of the holding interval 
is determined by the intensity of the transaction costs and the time horizon.
\end{abstract}

\section{Introduction}  
\label{s:introduction}
\vspace*{-0.5pt}
\noindent

An idealised model of investment is a sequence of gambles where the speculator
chooses at each time step her position. The game is multiplicative if the pay-off
is proportional to the capital, and it is Markov if the new capital
and new position depend parametrically only on the previous state.
The relevant issue consists of determining which strategy the speculator should 
pursue, which in general depends on the preferences of the individual investor.
A reasonable choice is to assume that the investor wishes
to maximise the growth of her capital. In this paper we will
investigate such strategies for finite-time investment horizons.
Another interpretation is then that we will be looking at investments
in a class of quadratic utility functions, where, for definiteness, we
take certain values of the parameters, which are then those
that correspond to growth-optimal investment. The main novelty is that
we include transaction costs. With the special functional form of the
costs studies here, the investment problem can be solved analytically
by mapping to an example in quantum mechanics.
 
Growth optimal criteria for multiplicative Markov process were first investigated
by Kelly in the context of information theory \cite{Kelly}. 
Criticisms from the viewpoint of utility theory
appeared later in the economic literature.
For recent reviews the reader is referred to refs. 
\cite{HakansonZiemba,MaslovZhang,AurellBavieraHammarlidServaVulpiani}. 
A related but separate question is what to do if the laws of
the gambles is unknown, but has to be deduced from observing 
price history (or other information). A widely recognised procedure
is then the ``Universal portfolios''\cite{Cover,CoverOrdentlich}, which can be considered
growth-optimal strategies, using continuously updated guesses for
the parameters of the model. Growth-optimal criteria are therefore 
robust to the market participnats lacking knowledge
of the statistical laws of the market. This makes them interesting also over finite-time
horizons. 

In a previous contribution \cite{AurellMG} we derived from a discrete 
multiplicative Markov model the continuum limit dynamics of a stock and 
bond portfolio in the presence of linear trading costs, i.e. proportional to 
the absolute value of the capital moved by the investor to balance her 
portfolio. Our main result can be summarised by saying that on an infinite
time horizon the investment optimal strategy consists of allowing the amount
of capital invested in stocks to fluctuate freely within an interval around 
the value of the optimal investment {\em in the absence} of trading costs. 
The size of the holding interval was shown to depend non-analytically 
on an adimensional parameter measuring the intensity of the transaction costs.
In \cite{AurellMG} we took the existence of growth-optimal strategies
over infinite time horizons
for granted. This has recently been proven, and in a much more general
setting \cite{EvstigneevSchenk-Hoppe}.  
A consequence of the result on the holding interval is that
that for financially reasonable values of the parameters in the model, convergence to the 
full dynamical solution in the infinite time horizon limit may be surprisingly 
slow, of the order of years of trading.
It is therefore natural to address the question of the time evolution of the
growth optimal strategy over finite investment horizons. The question has relevance also
in the perspective of deriving an option pricing procedure from growth optimal 
criteria. 

In the present paper we tackle the problem of an optimal 
strategy in a market model where trading costs are described by a 
quadratic function of the fraction of capital invested in re-hedging the 
portfolio. 
One advantage of this model that it is mathematically simpler, since the
costs are analytic. Indeed, we will show that the model can essentially
be solved analytically, which is interesting in itself. From the financial
side, one can compare with the market impact phenomenology of \cite{Farmer}.
Suppose one first buys $w$ worth of shares, and sells off shares to recover
the same amount $w$. By actively
seeking a deal one is forced to buy high and sell low. The turn-around 
cost of the operation is then on average $w^2/\lambda$, under the
assumption that $w$ is
much less than the market depth $\lambda$. 
If one assumes that market depth grows
proportionally to the total wealth of a typical investor in the market one recovers
the model we study here. The analysis we will
present can therefore, for example, be
relevant to fairly large operators in a market, 
the actions of which move market prices, to some extent.   

The paper is organised as follows. In section  \ref{sec:hjb} we state the optimisation
problem in the framework of the Hamilton-Jacobi-Bellman equation, 
well known in mathematical finance. In section \ref{sec:nocosts} we 
show that the non-linear Hamilton-Jacobi-Bellman equation governing the dynamics,
in our example,
is solvable in the small transaction costs limit by means of a multi-scale perturbation 
theory (see for example \cite{Fauve,ChenGoldenfeldOono}, or \cite{Frisch}, chapter 9). 
This is the main technical result of the paper, and reduces the non-linearity
to a normal form. All higher order corrections can be computed from ancillary 
linear non-homogeneous equations.
In section \ref{sec:logarithm} we solve analytically the normal form of the non-linearity
and compare the result with the numerical solution of the original Hamilton-Jacobi-Bellman
equation. The approximation turns out to be very accurate for realistic values of the 
parameters in the model. The last section is devoted to a discussion of the results.

\section{HJB equation for trading costs}
\label{sec:hjb}
\noindent
The stock and bond trade dynamics is in the continuum limit governed by the system
of stochastic differential equations
\begin{eqnarray}
&&dW_{t}=[\mu \rho_{t}-\gamma f^2(\rho_t,t)]\,W_{t} dt+\sigma \rho_t W_t \,dB_t
\label{hjb:wealthsde}\\
&&d\rho_t=\left[f(\rho_t,t)+a(\rho_t)+\gamma 
\rho_t f^2(\rho_t,t) \right]\,dt+b(\rho_t)\,dB_t
\label{hjb:stocksde}
\end{eqnarray}
with
\begin{eqnarray}
a(\rho_t)&=&\rho_t(1-\rho_t)(\mu-\sigma^2 \rho_t)
\nonumber\\
b(\rho_t)&=&\sigma \rho_t(1-\rho_t)
\label{hjb:drift}
\end{eqnarray}
In the equations (\ref{hjb:wealthsde})-(\ref{hjb:stocksde}), 
$W_t$ is the total wealth of the speculator at time $t$, and $\rho_t$ is
the fraction of the wealth held in stocks at time $t$. 
The stochastic control $f$ represents the action taken by the speculator at time $t$ to 
re-hedge her position in the market. The stochastic control is to be
determined as a function of
$\rho_t$ and $t$, by maximising the expectation value of the wealth growth:
\begin{eqnarray}
\lambda(x,t;T)=E_{\rho_t=x}\ln \frac{W_T}{W_t}=E_{\rho_t=x}\int_{t}^{T}ds\left[\mu \rho_s-
\frac{\sigma^2}{2}\rho_s^2-\gamma f^2\right]
\label{hjb:growth}
\end{eqnarray}
The expectation $E_{\rho_t=x}$ is conditional on the fraction in stock process $\rho_t$ 
having value $x$ at initial time $t$. The time difference $T-t$ is the time horizon of the speculator: 
the time period wherein she wants to optimise her position in the market.  
The optimisation is performed with respect to two conflicting effects.
On one hand, the market fall-outs raise or lower the relative amount of invested wealth,
motivating the investor to re-balance the portfolio. On the other, 
the re-hedging carries trading costs.
In this paper we model these as $\gamma f^2$, where $\gamma$ is some given positive 
valued constant.
The derivation of the equations (\ref{hjb:wealthsde})-(\ref{hjb:stocksde}) from a discrete 
multiplicative Markov game is given in appendix.

For any stochastic control $f$ such that the system (\ref{hjb:wealthsde}), (\ref{hjb:stocksde})
is well defined, the expectation value of the wealth growth must obey the dynamic programming
equation
\begin{eqnarray}
&&\partial_{t}\lambda+[f+ a+\gamma x f^2]\partial_{x}\lambda+\frac{b^2}{2}
\partial_{x}^{2}\lambda+\mu x -\frac{\sigma^2 x^{2}}{2}-\gamma f^2=0
\nonumber\\
&&\lambda(x,T;T)=0 
\label{hjb:dynamicprogramming}
\end{eqnarray}
The functional dependence of the drift $a$ and of the diffusion coefficient $b$ on $x$ 
is defined by (\ref{hjb:drift}) and has been omitted to streamline the notation.
The optimisation problem is well defined only when the boundary conditions in $x$ are 
specified.
The solution of the dynamic programming equation (\ref{hjb:dynamicprogramming}) has by 
construction the form of an average over the transition probability density of the 
stochastic process $\rho_t$. 
In the model considered here we can neglect capital borrowing and lending
since (under proper conditions on the parameters) the process will never move to 
$\rho$ outside the interval $[0,1]$. The conservation of the probability measure 
then implies
\begin{eqnarray}
\de_{x} \lambda(x,t;T)|_{x=0}= \de_{x} \lambda(x,t;T)|_{x=1}=0
\label{hjb:reflecting}
\end{eqnarray}
which are adjoint to the reflecting boundary conditions imposed on the probability 
density.
  
The growth is stationary versus the control $f$ if
\begin{eqnarray}
\frac{\delta \lambda(.,t)}{\delta f(.,t')}=0
\end{eqnarray}
For any finite $\gamma$, the extremum condition yields a relation between 
the stochastic control and the expected growth:
\begin{eqnarray}
(1+2\,\gamma x f)\partial_{x}\lambda-2\,\gamma\,f=0
\end{eqnarray}
The extremum is a maximum for
\begin{eqnarray}
\Phi(f,\lambda)=[f+ a+\gamma x f^2]\partial_{x}\lambda+\frac{b^2}{2}
\partial_{x}^{2}\lambda+\mu x -\frac{\sigma^2 x^{2}}{2}-\gamma f^2
\end{eqnarray}
provided \cite{Oksendal,FlemingMeteSoner}
\begin{eqnarray}
2\,\gamma \,x\,\partial_{x}\lambda-2\,\gamma<0
\end{eqnarray}
Thus, the optimal stochastic control is
\begin{eqnarray}
f=\frac{\partial_{x}\lambda}{2\,\gamma\,(1-x\,\partial_{x}\lambda)}
\end{eqnarray}
The Hamilton-Jacobi-Bellman \cite{Oksendal} equation governing the optimal dynamics
is then
\begin{equation}
\begin{array}{l}
\partial_{t}\lambda+a\,\partial_{x}\lambda+
\frac{(\de_x\lambda)^2}{4\,\gamma\,(1-x\,\de_x\lambda)}+
\frac{b^2 }{2}\partial_{x}^{2}\lambda+\mu x-\frac{\sigma^2 x^{2}}{2}=0 \\[2mm]
\lambda(x,T;T)=0 \\[2mm]
\de_{x} \lambda(x,t;T)|_{x=0}= \de_{x} \lambda(x,t;T)|_{x=1}=0\\
\end{array}
\label{hjb:hjb}
\end{equation}
This equation is {\em time autonomous} and therefore the solution can be sought in the
form
\begin{eqnarray}
\lambda(x,t;T)\equiv\lambda(x,T-t)
\end{eqnarray}
In view of the ensuing analysis of the Hamilton-Jacobi-Bellman equation
it is useful to identify the canonical dimensions of the quantities involved 
in the problem:
\begin{equation}
\begin{array}{lclclcl}
  [\lambda] &=&        0&\qquad\qquad &[x]   &=&0 \\[2mm] 
  [\sigma^2]&=&    [1/t]&\qquad\qquad &[\mu] &=&[1/t]\\[2mm] 
  [\gamma]  &=&      [t]&\qquad\qquad &[f]   &=&[1/t]\\
\end{array}
\label{hjb:dimensions}
\end{equation}
A trading day can be assumed to define the time unit.
Note that the dimensions in (\ref{hjb:dimensions}) are partially different from
those given in \cite{AurellMG} since the functional form of the
friction is different.
\section{The transaction cost free limit and its leading order correction}
\label{sec:nocosts}
\noindent
In the absence of transaction costs the speculator is free to take un-restrained actions
to always keep the fraction allocated to stocks constant
\begin{eqnarray}
\rho^{\mathrm{opt}}=\frac{\mu}{\sigma^2}
\label{nocosts:idealoptimum}
\end{eqnarray}
This gives \cite{Aase,Oksendal} the absolute value of the wealth growth:
\begin{eqnarray}
\lambda(x,t;T)|_{\gamma=0}=\frac{\mu^2}{2\,\sigma^2}(T-t)
\label{nocosts:idealgrowth}
\end{eqnarray}
In this paper we assume $\mu$ and $\sigma^2$ are such that
the ratio (\ref{nocosts:idealoptimum}) lies inside the
interval $[0,1]$, see the appendix for discussion.

In the frame-work of the dynamics programming equation, the absence of trading costs
renders (\ref{hjb:dynamicprogramming}) linear in the stochastic control $f$. The optimal
growth is achieved by wielding a singular control strategy which restricts the support of the 
probability measure of $\rho_t$ only to the point $\rho^{\mathrm{opt}}$.
Thus, when trading costs are present but ``small'', it must be possible to seek to 
solve of the Hamilton-Jacobi-Bellman equation (\ref{hjb:hjb}) by means of a perturbative
expansion around the limit (\ref{nocosts:idealgrowth}). In order to do that it is convenient
to translate the origin of the $x$ coordinate according to 
\begin{eqnarray}
x\,\rightarrow\,x+\frac{\mu}{\sigma^2}
\end{eqnarray}
For any finite $\sigma^2$, the adimensional parameter measuring the intensity of 
trading costs is
\begin{eqnarray}
\epsilon=\sigma^2\,\gamma
\label{nocosts:epsilon}
\end{eqnarray}
However, the perturbative expansion cannot be analytic in $\epsilon$.
This can be argued a priori by observing that Oseledec' theorem 
\cite{Oseledec} predicts for the asymptotic behaviour of the wealth growth
\begin{eqnarray}
\lim_{t\,\downarrow\,-\infty}\lambda(x,t;T)\sim (T-t)\,\ell
\label{nocosts:Oseledec}
\end{eqnarray}
It is therefore natural to assume that the dynamics of the wealth growth $\lambda$
should involve two typical time scales. The first time scale should be 
$\ell^{-1}$ governing the asymptotic regime, while the second should describe the 
characteristic relaxation time to the asymptotic regime. Such considerations, together with 
the singular dependence of the Hamilton-Jacobi-Bellman equation (\ref{hjb:hjb}) 
on $\gamma$, suggest the $Ansatz$
\begin{eqnarray}
\lambda(x,t;T)=\epsilon\, \varphi\left(\frac{x}{\epsilon^{1/4}},\frac{T-t}{\epsilon},
\frac{T-t}{\epsilon^{1/2}}\right)
\label{nocosts:ansatz}
\end{eqnarray}
The first time dependence refers to the linear growth. The second describes the 
relaxation process and it is conjectured to satisfy in $\epsilon$ the characteristic 
time-spatial scaling relation of diffusion processes. From (\ref{nocosts:ansatz})
it appears that small transaction cost expansion of $\lambda$ entails a multi-scale 
perturbation theory with times
\begin{eqnarray}
&&r:=\frac{T-t}{\epsilon}
\nonumber\\
&&s:=\frac{T-t}{\epsilon^{1/2}}
\label{nocosts:twoscale}
\end{eqnarray}
and consequently
\begin{eqnarray}
\de_{t}=\frac{1}{\epsilon}\de_r+\frac{1}{\epsilon^{1/2}}\de_s
\end{eqnarray}
The ``spatial'' rescaling
\begin{eqnarray}
y:=\frac{x}{\epsilon^{1/4}}
\end{eqnarray}
finally yields for
\begin{eqnarray}
\varphi=\varphi(y,r,s)
\end{eqnarray}
the equation
\begin{eqnarray}
0&=&\de_r \varphi+\frac{\mu^2}{2\,\sigma^2}+
\epsilon^{1/2} \left[\de_s \varphi+\frac{\sigma^4 (\de_y\varphi)^2}{ 
4(\sigma^2+\epsilon^{3/4} (\mu + \epsilon^{1/4} y \sigma^2) \de_{y}\varphi)}
\right.
\nonumber\\ &&\left.
+\,\frac{(\mu + \epsilon^{1/4} y \sigma^2)^2 (\mu + (\epsilon^{1/4}\, y-1)\sigma^2)^2}{2\,\sigma^6} 
\de_{y}^{2}\varphi  -\frac{\sigma^2 y^2}{2}\right]
\nonumber\\
&&+\,\frac{\epsilon\, y (\mu + \epsilon^{1/4} y\ \sigma^2)(\mu + (\epsilon^{1/4}\, y-1)\sigma^2)}{\sigma^2}
\de_{y}\varphi
\end{eqnarray}
The equation admits now an analytic expansion in powers of $\epsilon^{1/4}$ 
the solution whereof is amenable to the form of the series
\begin{eqnarray}
\varphi=\frac{\mu^2\,r}{2\,\sigma^2}+\sum_{n=0}^{\infty} \epsilon^{n/4}\,\varphi_{n}(y,s)
\label{nocosts:expansion}
\end{eqnarray}
The equation for $\varphi_{0}(y,s)$ is non-linear and provides the {\em normal form}
of the non-linearity involved in the Hamilton-Jacobi-Bellman problem (\ref{hjb:hjb}):
\begin{eqnarray}
\partial_{s}\varphi_0+\frac{\sigma^2\,(\de_x\varphi_0)^2}{4}+
\frac{D^2}{2}\partial_{x}^{2}\varphi_0-\frac{\sigma^2 y^{2}}{2}=0
\label{nocosts:normal}
\end{eqnarray}
The effective ``diffusion'' constant $D^2$ in (\ref{nocosts:normal}) is
\begin{eqnarray}
D^2=\sigma^2\left(\frac{\mu}{\sigma^2}\right)^2\left(1-\frac{\mu}{\sigma^2}\right)^2\,,
\qquad [D^2]=[1/t] 
\end{eqnarray}
Terms of higher orders in the expansion (\ref{nocosts:expansion})
are obtained by solving linear non-homogeneous equations. The first of them 
is for example provided by the solution of
\begin{eqnarray}
&&\de_s\varphi_1+\frac{1}{2}\sigma^2 \de_{y}\varphi_{0} \de_{y}\varphi_{1} + 
\frac{D^2}{2}\de_y^2\varphi_1
\nonumber\\
&&\quad +\,\frac{(2\,\mu^2\,y \sigma^2(\mu - \sigma^2) + 2 
\mu y \sigma^2 (\mu - \sigma^2)^2)}{2\,\sigma^6}\de_y^2\varphi_{0}=0
\end{eqnarray}
The conclusion of the above analysis is that the leading order in the multi-scale 
expansion captures the effect of the non-linearity involved in the optimisation
problem. Otherwise stated, within leading order in the small trading costs limit, 
equation (\ref{hjb:hjb}) can be consistently replaced by the simpler {\em model problem} 
\begin{eqnarray}
&&\partial_{t}\lambda+\frac{(\de_x\lambda)^2}{4\,\gamma}+
\frac{D^2}{2}\partial_{x}^{2}\lambda+\frac{\mu^2}{2\,\sigma^2}-\frac{\sigma^2 x^{2}}{2}=0
\nonumber\\
&&\lambda(y,T;T)=0
\nonumber\\
&&\de_{x} \lambda(x,t;T)\left|_{x=-\frac{\mu}{\epsilon^{1/4}\,\sigma^2}}\right.= 
\de_{x}\lambda(x,t;T)\left|_{x=\frac{1}{\epsilon^{1/4}}\left(1-\frac{\mu}{\sigma^2}\right)}
\right.=0
\label{nocosts:model}
\end{eqnarray}  
The model problem is the Hamilton-Jacobi-Bellman equation associated to the 
dynamic programming equation
\begin{eqnarray}
&&\partial_{t}\lambda+\sup_{f}\,\{f\,\partial_{x}\lambda+
\frac{D^{2}}{2}\,\partial_{x}^{2}\lambda 
+\frac{\mu^2}{2\,\sigma^2} -\frac{\sigma^2 x^{2}}{2}-\gamma f^2\}=0
\label{nocosts:dp}
\end{eqnarray}
The extremum conditions here become
\begin{eqnarray}
&&\de_{x}\lambda-2\,\gamma f=0
\nonumber\\
&&-2\,\gamma<0
\end{eqnarray}
stating that the solution is always a maximum. This fact guarantees that (\ref{nocosts:model}) 
provides the optimal capital growth for $\epsilon$ small enough.

\section{The logarithmic transform}
\label{sec:logarithm}
\noindent
The model problem (\ref{nocosts:model}) can be mapped to 
a linear equation by the logarithmic 
transform
\begin{eqnarray}
\lambda(x,t;T)=A \ln \psi(x,T-t)
\label{logarithm:log}
\end{eqnarray}
provided 
\begin{eqnarray}
\psi(x,T-t)\,>\,0
\end{eqnarray}
The initial condition for the function $\psi$ is
\begin{eqnarray}
\psi(x,0)\,=\,1
\end{eqnarray}
The value of the constant $A$ is fixed by imposing the cancellation of the non-linear 
term. Namely, by setting
\begin{eqnarray}
A=2\,D^2\,\gamma\,,\qquad [A]=0
\end{eqnarray}
(\ref{nocosts:dp}) is mapped into the imaginary time Schr\"odinger
equation 
\begin{eqnarray}
-\de_t\psi+\frac{D^2}{2}\de_{x}^2\psi
+\frac{1}{2\,D^2\,\gamma}\left[\frac{\mu^2}{2\,\sigma^2}-\frac{\sigma^2\,x^2}{2}\right]\psi=0
\label{logarithm:ho}
\end{eqnarray}
The equation further simplifies if the Neumann boundary conditions on a finite interval of
length proportional to $\epsilon^{-1/4}$ are replaced by Dirichlet boundary conditions at
infinity:
\begin{eqnarray}
\lim_{|x|\,\uparrow\,\infty}\psi(x,t)=0
\end{eqnarray}
As a matter of fact, (\ref{logarithm:ho}) admits on ${\mathbb{L}}^{2}(\mathbb{R})$ the 
explicit solution (see for example appendix~A.4 in \cite{Sakurai}):
\begin{eqnarray}
\psi(x,t)=e^{\frac{1}{2\,D^2\,\gamma}\frac{\mu^2\,t}{2\,\sigma^2}}\sum_{n=0} 
e^{-\,E_{2 n} t}\,\psi_{2 n}(x)\int_{\mathbb{R}}dy\,\psi_{2 n}(y)
\label{logarithm:solution}
\end{eqnarray}
where
\begin{eqnarray}
&&\psi_n(x)=\frac{1}{\sqrt{2^n\,n!}}\left(\frac{1}{\pi\,\tau\, D^2}\right)^{1/4}\,
e^{-\frac{x^2}{2\,\tau D^2}}\,H_{n}\left(\frac{x}{\sqrt{\tau\,D^2}}\right)
\nonumber\\
&&E_n=\frac{1}{\tau}\left(n+\frac{1}{2}\right)\,,\qquad\qquad \frac{1}{\tau}=
\sqrt{\frac{\sigma^2}{2\,\gamma}}
\end{eqnarray}
for $H_{n}(\xi)$ the $n$-th Hermite polynomial. The series (\ref{logarithm:solution}) 
is restricted to even values of $n$ since Hermite polynomials are
of even/odd parity for even/odd $n$ 
\begin{eqnarray}
H_{n}(-\xi)=(-)^{n} H_{n}(\xi)
\end{eqnarray}
The logarithmic transform is well defined on (\ref{logarithm:solution}) 
due to the exponential decay of the coefficients in the series
\begin{eqnarray}
\frac{1}{\sqrt{2^n\,n!}}
\left(\frac{1}{\pi\,\tau\, D^2}\right)^{1/4}
\int_{\mathbb{R}}dy\,\psi_{2 n}(y)=\frac{ (2\,n-1)!! \sqrt{2}}{2^n (2\,n)!}
\end{eqnarray}
Hence, the average optimal growth of the capital as a function of the fraction initially
invested in stocks $x$ on a time horizon $T-t$ is
\begin{eqnarray}
&&\lambda(x,t;T)=\left[\frac{\mu^2}{2\,\sigma^2}-\frac{D^2\,\sigma^2\,\tau}{2}
\right]\,(T-t)- \frac{\sigma^2\,\tau\,x^2}{2}+ D^2\,\sigma^2\,\tau^2\, \ln \sqrt{2}
\nonumber\\
&&\qquad +\,D^2\,\sigma^2\,\tau^2\, \ln 
\left\{\sum_{n=0}^{\infty} \frac{(2\,n-1)!!\,e^{-\,2\, n\,\frac{T-t}{\tau}}}{2^n (2\,n)!}\,
H_{2\,n}\left(\frac{x}{\sqrt{\tau\,D^2}}\right)\right\}
\label{logarithm:sol}
\end{eqnarray}
In agreement with Oseledec' theorem (\ref{nocosts:Oseledec}), 
convergence to a stationary state is exponentially fast with rate equal to 
\begin{eqnarray}
\frac{\tau}{2}=\frac{1}{2}\sqrt{\frac{2\,\gamma}{\sigma^2}}\equiv\frac{1}{\sigma^2}
\sqrt{\frac{\epsilon}{2}}
\label{logarithm:dacay}
\end{eqnarray}
\begin{figure}
\begin{center}
\mbox{\includegraphics[height=6cm,width=10cm]{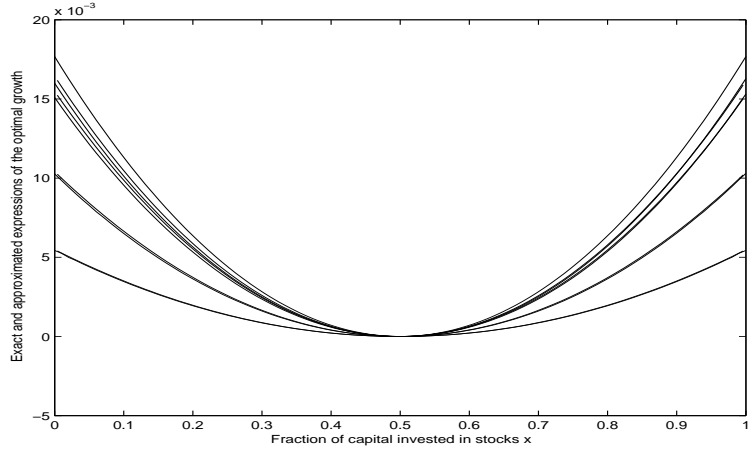}}
\end{center}
\caption{The control potential $2\,\gamma\,V$ as defined in (\protect\ref{discussion:potential})
for $\mu=\sigma^2/2$, $\sigma=10^{-2}$ and $\gamma=10^{2}$ is plotted for time horizons
of $T-t=500,1000,2000,2500$ days using both the exact solution of the Hamilton-Jacobi-Bellman 
equation (\protect\ref{hjb:hjb}) and the approximate solution (\protect\ref{logarithm:sol}).
The characteristic decay time to the asymptotic regime is $\tau/2\sim1000$ days. The 
innermost parabola is obtained from the asymptotic expression 
(\protect\ref{discussion:asymptotic}).}
\label{fig:potential}
\end{figure}
From the last identity it is also straightforward to verify the consistency of (\ref{logarithm:sol})
with the multi-scale Ansatz (\ref{nocosts:ansatz}).
Finally, the time asymptotic form of the solution can be extracted directly from 
(\ref{nocosts:model}) as it was done in our previous paper \cite{AurellMG} by using
the information (\ref{nocosts:Oseledec}) provided by Oseledec' theorem.

\section{Discussion and conclusions}
\noindent
The exact solution of the model problem (\ref{nocosts:model}) gives a qualitatively
and for $\sigma^2\,\gamma$ small enough also quantitatively correct description of the 
investment strategy that a speculator should pursue in order to optimise her profits.
The strategy is most conveniently summarised by looking at the potential $V$
\begin{eqnarray}
f(x,T-t)=-\de_x V(x,T-t) 
\end{eqnarray}
associated to the optimal stochastic control.
Within the same approximation leading to (\ref{nocosts:model}), the potential
can be defined as
\begin{eqnarray}
V(x,T-t)=-\frac{\lambda(x,T-t)}{2\,\gamma}+ \frac{\lambda(0,T-t)}{2\,\gamma}
\label{discussion:potential}
\end{eqnarray}
The behaviour in time of the potential is illustrated in figure~\ref{fig:potential}. In the figure
the same quantity (\ref{discussion:potential}) is also plotted when $\lambda$ is obtained from 
the solution of the exact Hamilton-Jacobi-Bellman equation (\ref{hjb:hjb}). The resulting profiles 
are practically indistinguishable for financially reasonable choices of the parameters in the
perturbative regime. 
The shape of the potential entails a fast decay of the probability density of the fraction
of capital invested in stocks versus the deviation from the ideal optimum 
(\ref{nocosts:idealoptimum}). The observation justifies a posteriori the use of Dirichlet 
boundary conditions in the solution of the model problem. As a matter of fact, the modification 
of the boundary conditions affects the average (\ref{hjb:growth}) only in a region where the 
probability density is practically equal to zero.
\begin{figure}
\begin{center}
\mbox{\includegraphics[height=6cm,width=10cm]{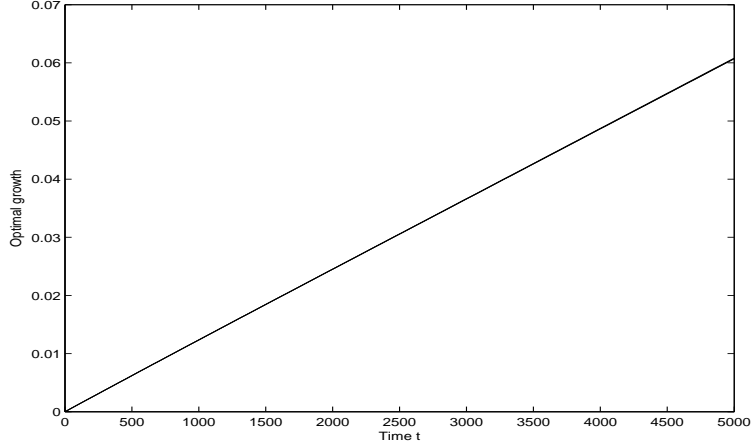}}
\end{center}
\caption{The exact and approximate optimal growth $\lambda(0,t)$ is plotted versus
the time horizon for the same parameters used in (\ref{discussion:asymptotic}). 
The two lines almost overlap.}
\label{fig:growth}
\end{figure}

The potential gets steeper when time to investment horizon
is long, tending asymptotically to a parabolic shape 
\begin{eqnarray}
V_{\mathrm{asympt.}}(x)=\frac{x^2}{2\tau}\qquad(T-t) >> \tau
\label{discussion:asymptotic}
\end{eqnarray}
In this limit the speculator aims to always
hold the invested 
fraction of capital in a finite interval around the optimal 
investment fraction $\rho^{\mathrm{opt}}$ of (\ref{nocosts:idealoptimum}). 
In the asymptotic regime the fraction of capital 
invested in stocks tends to an Ornstein-Uhlenbeck process \cite{BorodinSalminen} the invariant
measure whereof having variance $D^2 \tau/2$. This latter quantity provides the typical size of 
the holding interval. When $T-t$ is on the order of $\tau$, or smaller, the solution
is given by the complete expression (\ref{logarithm:sol}), with corrections from
the higher order terms in (\ref{nocosts:expansion}). The general behaviour of this
process is a successively flatter potential leading to a larger holding interval, see
Fig.~\ref{fig:growth}.

The results of the present paper together with those of our previous contribution \cite{AurellMG} 
support an ``investment confinement'' picture as growth optimal strategy for multiplicative 
Markov market models with trading costs. 
According to such picture, differences in the modeling of the trading costs 
are reflected only in the different non-analytic powers of $\epsilon$ on which the size 
of the holding interval depends. Differences in the modeling of transaction costs do not 
seem to affect the typical time scales governing the relaxation to the asymptotic growth state.
Namely also in the case of quadratic market friction, a daily relative stock price fluctuation 
$\sigma^2$ of the order of $10^{-4}$ per cent with $\epsilon$ equal to $0.01$ yields for 
decay rate the value of $700$ days. The result is in agreement with the prediction
of the dimensional analysis in \cite{AurellMG}.

\section*{Acknowledgements}
\noindent
The authors gratefully acknowledge discussions with A.~Kupiainen. 
This work was supported by the Swedish Research Council through
grant NFR I~510-930 (E.A.), and 
grant ``hanke'' nr. 39746 `` p\"a\"at\"os'' nr. 
68820 from the Finnish Academy of Science (P.M.G.).

\section*{Appendix}

We present here a derivation of the continuum limit market dynamics slightly
different than the one given in~\cite{AurellMG}.
At time $t$ the wealth in stocks is, in units of the total wealth,
\begin{equation}
W^{(Stocks)}_t=\rho_t\,W_t
\label{app:wealthinstocks}
\end{equation}
The variation in one time step of the wealth in stocks occurs in consequence of
\begin{itemize}
\item the market fall-out $u_t$
\item the action $\Delta \chi_t$ of the speculator who re-hedges her position in the market.
\end{itemize}
The fraction in stocks at time $t+1$ becomes
\begin{equation}
W^{(Stocks)}_{t+1}=\left[u_t \rho_t +\Delta \chi_t\right]\, W_t
\label{app:stockincrement}
\end{equation}
The total wealth at time $t+1$ is affected by the stock investment profits 
or losses and by the trading costs entailed by any re-hedging:
\begin{equation}
W_{t+1}=\left[1+\rho_t (u_t -1) -\Delta F_{\gamma}(\Delta \chi_t)\right]\, W_t
\label{app:wealthincrement}
\end{equation}
Most generally trading costs are described by a semi-positive definite
function $\Delta F_{\gamma}$ vanishing only if the investor 
remains idle, i.e. when $\Delta \chi$ is zero.\\
From (\ref{app:stockincrement}), (\ref{app:wealthincrement}) the variation 
of the invested capital fraction $\rho_t$ over a time unit is 
\begin{equation}
\Delta \rho_t=\frac{\rho_t+(u_t-1)\rho_t+ \Delta \chi_t}{1+\rho_t (u_t-1)-
\Delta F_{\gamma}(\Delta \chi_t)}-\rho_t
\label{app:rhochange}
\end{equation}
The continuum limit is attained by replacing
\begin{eqnarray}
&& u_t-1\,\rightarrow\,\mu\,dt+\sigma\,dB_t
\nonumber\\
&& \Delta\chi_t\,\rightarrow\,f\,dt
\nonumber\\
&& \Delta F\,\rightarrow\,\gamma {\mathcal{F}}(f)dt
\end{eqnarray}
The differential $du_t$ gives the relative stock price
\begin{eqnarray}
du_t:=\frac{d p_t}{p_t}=\mu\,dt+\sigma\,dB_t
\end{eqnarray}
The stochastic differential equation is defined according to the Ito convention.
It has the solution
\begin{eqnarray}
p_t=p_o\,e^{\left(\mu-\frac{\sigma^2}{2}\right)\,t+\sigma\,B_t}
\end{eqnarray}
A value of $\frac{\mu}{\sigma^2}$ outside the interval $[0,1]$ thus
corresponds to strong inflation or deflation rates.
If borrowing and short-selling is not allowed,
the optimal strategy would then simply be to keep all money in stock or all
money in bonds. If borrowing and short-selling is allowed,
the problem becomes again similar to the one studied here, but the
relevant intervals would then either be $[1,\infty]$ or $[-\infty,0]$.
   
After a little algebra one finds
\begin{eqnarray}
&&dW_{t}=\mu \rho_{t}\,W_{t} dt+\sigma \rho_t W_t \,dB_t 
-\gamma W_{t}{\mathcal{F}} dt
\label{app:wealthrhosde}\\
&&d\rho_t=\left[f+\rho_t(1-\rho_t)(\mu-\sigma^2 \rho_t)\right]\,dt+\left[\sigma \rho_t(1-\rho_t)\right]\,dB_t+
\gamma\,\rho_t {\mathcal{F}}dt
\label{app:rhosde}
\end{eqnarray}
Equations (\ref{hjb:wealthsde}) and (\ref{hjb:stocksde}) are recovered by setting
\begin{eqnarray}
{\mathcal{F}} = f^2
\end{eqnarray}
The optimal control diverges in the limit of zero transaction costs so to
hold the fraction in stocks tightly to the constant value $\mu/\sigma^2$.


\end{document}